\documentclass[twocolumn]{aastex631}

\usepackage{amsmath,esint}

\usepackage{lipsum}
\usepackage{enumitem}
\usepackage{graphicx}
\usepackage{graphbox}

\begin{document}
\title{THE EFFECT OF POLOIDAL MAGNETIC FIELD AND HELICITY INJECTION IN A BREAKOUT CME \footnote{Released on March, 1st, 2021}}

\correspondingauthor{Dipankar Banerjee}
\email{dipubando@gmail.com, dipu@iist.ac.in}

\author[0000-0002-0786-7307]{Nitin Vashishtha}
\affiliation{Aryabhatta Research Institute of Observational Sciences, Beluwakhan, Nainital, India}
\affiliation{Department of Physics, Deen Dayal Upadhyaya Gorakhpur University, Gorakhpur 273009, India}
\author[0000-0002-6954-2276]{Vaibhav Pant}
\affiliation{Aryabhatta Research Institute of Observational Sciences, Beluwakhan, Nainital, India}

\author[0000-0002-9311-9021]{Dana-Camelia Talpeanu}
\affiliation{Solar-Terrestrial Centre of Excellence - SIDC, Royal Observatory of Belgium, Avenue Circulaire 3, Brussels 1180, Belgium}
\author[0000-0003-4653-6823]{Dipankar Banerjee}
\affiliation{Aryabhatta Research Institute of Observational Sciences, Beluwakhan, Nainital, India}
\affiliation{Indian Institute of Astrophysics, 2nd Block Koramangala, Bangalore, India}
\affiliation{Center of Excellence in Space Science, IISER Kolkata, Kolkata, India}
\author[0000-0002-2397-5199]{Shantanu Rastogi}
\affiliation{Department of Physics, Deen Dayal Upadhyaya Gorakhpur University, Gorakhpur 273009, India}





\begin{abstract}

Coronal mass ejections (CMEs), as crucial drivers of space weather, necessitate a comprehensive understanding of their initiation and evolution in the solar corona, in order to better predict their propagation. Solar Cycle 24 exhibited lower sunspot numbers compared to Solar Cycle 23, along with a decrease in the heliospheric magnetic pressure. Consequently, a higher frequency of weak CMEs was observed during Solar Cycle 24. Forecasting CMEs is vital, and various methods, primarily involving the study of the global magnetic parameters using datasets like Space-weather Helioseismic and Magnetic Imager Active Region Patches (SHARP), have been employed in earlier works. In this study, we perform numerical simulations of CMEs within a magnetohydrodynamics framework using Message Passing Interface - Adaptive Mesh Refinement Versatile Advection Code (MPI-AMRVAC) in 2.5 dimensions. By employing the breakout model for CME initiation, we introduce a multipolar magnetic field configuration within a background bipolar magnetic field, inducing shear to trigger the CME eruption. Our investigation focuses on understanding the impact of the background global magnetic field on CME eruptions. Furthermore, we analyze the evolution of various global magnetic parameters in distinct scenarios (failed eruption, single eruption, multiple eruptions) resulting from varying amounts of helicity injection in the form of shear at the base of the magnetic arcade system. Our findings reveal that an increase in the strength of the background poloidal magnetic field constrains CME eruptions. Furthermore, we establish that the growth rate of absolute net current helicity is the crucial factor that determines the likelihood of CME eruptions.

\end{abstract}



\section{Introduction} \label{sec:intro}
Coronal mass ejections (CMEs) are some of the significant and dynamic phenomena that involve the ejection of a massive amount of plasma and magnetic field from the Sun's corona into the heliosphere \citep{hundhausen1984, Gopalswamy2004, yashiro2004, Webb2012}. CMEs are acknowledged as key catalysts of space weather disturbances on Earth, given their capacity to initiate geomagnetic storms, potentially leading to technological disruptions on our planet \citep{hapgood, Schrijver2015, Schrijver_etal2015}. They are capable of moving at velocities spanning from a few hundred to several thousand km s$^{-1}$, and can undergo acceleration ranging from a few tens to a few 10$^{4}$ m s$^{-2}$ \citep{Webb2012}. The frequency of CME occurrences varies with the solar cycle in terms of amplitude and phase \citep{howard1986solar}. This frequency spans from one CME daily during solar minimum to nearly five CMEs daily during solar maximum \citep{st1999comparison,gopalswamy2005coronal,gopalswamy2006pre,Webb2012}, with approximately 4\% being directed along the Sun-Earth line \citep{Gopalswamy2004}. \cite{gopalswamy2010coronal} reported that halo CMEs, including those Earth-directed, are faster and more energetic. Understanding the mechanisms that initiate and propagate CMEs is crucial for comprehending space weather occurrences and enhancing prediction capabilities. In this context, numerical magnetohydrodynamic (MHD) modelling has emerged as a valuable method for offering a comprehensive overview of Sun-Earth interactions, from the initiation of a CME in the Sun's lower corona to its propagation to Earth.\par
However, a unanimous agreement regarding the precise triggering mechanism of CMEs is lacking. Various models suggest that the primary energy responsible for CMEs must originate from the coronal magnetic field \citep{chen1996, Chen2011}. For the pre-CME magnetic configuration, certain models \citep{van_ball, Forbes1991, Wu1997,gibson1998,krall2000, roussev2003,roussev2004} propose that a magnetic flux rope positioned above the solar surface is destabilized due to factors like footpoints motion, injection of magnetic helicity, or the draining of the filament material. On the other hand, alternative models \citep{mikic1988,mikic1994, Antiochos_1998, Antiochos_1999, Amari_2003a, Amari_2003b} start with the sheared magnetic structure that becomes unstable through the reconnection process \citep{gosling1993}. Specifically, the model proposed by \cite{Antiochos_1999} (Breakout Model) involves the multipolar magnetic field topology, in which the magnetic reconnection process occurs between the sheared arcade and the overlying magnetic field, creating a passage for the CME. \par
Interestingly, Solar Cycle 24 exhibited an unexpectedly high rate of CMEs per sunspot number, prompting significant discussion regarding its underlying causes \citep{LUHMANN2013221, Gopalswamy2014, wang2014ApJ...784L..27W,  petrie2015ApJ...812...74P,hess2017ApJ...836..134H, michalek2019ApJ...880...51M, Gopalswamy2020}. Solar Cycle 24 was preceded by an unusually prolonged minimum characterized by exceptionally low solar activity \citep{rusell2010RvGeo..48.2004R, Jiang2015}, a significantly weakened polar magnetic field \citep{munoz2012ApJ...753..146M}, and a reduced level of heliospheric open magnetic flux \citep{smith2008GeoRL..3522103S, owen2013LRSP...10....5O}. \cite{wang2014ApJ...784L..27W} and \cite{hess2017ApJ...836..134H} argued that the apparent increase in CME rates could result from observational biases, such as the doubling of the image cadence of the Large Angle and Spectrometric Coronagraph \citep[LASCO;][]{lasco1995SoPh..162..357B} on board the Solar and Heliospheric Observatory \citep[SOHO;][]{soho1995SoPh..162....1D} after 2010, potentially inflating the detection of smaller or fainter events. \cite{wang2014ApJ...784L..27W} argued that any influence of the polar fields on the global CME rate is likely a second-order effect. However, later studies, including those by \cite{Petrie_2013, petrie2015ApJ...812...74P}, \cite{Gopalswamy2015} and \cite{michalek2019ApJ...880...51M} supported a physical origin for the enhanced rate of weaker CMEs. These studies link the increase in the CME rate to a substantial decline in heliospheric and coronal total pressure \citep{mccomas2013ApJ...779....2M, Gopalswamy2015}, and a ~40\% reduction in the polar magnetic field strength of the Sun following the Cycle 23 reversal \citep{LUHMANN2013221, Petrie_2013}.
\par
Nonetheless, the challenge of predicting CMEs is amplified by the absence of a single physical pathway leading to the launch of the CME. Various attributes of active regions (ARs) have been recognized as precursors to flares and CMEs, including high magnetic field gradients, strong shear along polarity inversion lines, flux cancellation, and more \citep[][and references therein]{Toriumi2019}. Extensive efforts have been dedicated to study the progression of global magnetic parameters and the infusion of magnetic energy and helicities within active regions \citep{Zhang2001, Pariat2017, Zuccarello2018, Moraitis2021, Li2021, Liokati2022, Liu2023}. These global magnetic parameters turned out to be good indicators for the prediction of these eruptive events \citep{bobra_and_couvidat2014, Sinha2022, raju}. For predicting flares and flare-associated CMEs based on active regions \citep{bobra_and_couvidat2014, Sinha2022, raju}, researchers are utilizing the Space-weather HMI Active Region Patch (SHARP) data products \citep{bobra2014}. These products offer automated tracking of magnetic flux concentrations on the solar disk. Several of these studies have employed the SHARP data products to compute various parameters, which include metrics such as total unsigned current helicity, total Lorentz force magnitude, absolute net current helicity, and more. With the progression of machine learning methods, these predictions are continuously improving over time. Therefore, to comprehend the underlying physical processes of these eruptive occurrences and to pinpoint potential indicators for space weather forecasting tools, we conducted numerical MHD simulations using the breakout CME model \citep{Antiochos_1999}. Conducting numerical simulations is essential for gaining deeper insights into the initiation and propagation mechanisms of these phenomena. These insights will also enhance our ability to predict these impactful events, which have the potential to induce geomagnetic storms and interfere with satellite and telecommunication systems \citep{hapgood, Schrijver2015, Schrijver_etal2015}. Numerical simulations allow us to examine the temporal progression of magnetic parameters across varying levels of helicity injection, leading to both CMEs and failed eruptions. We also investigate the effect of the global coronal field on the initiation of the CMEs 
 with numerical simulations.
\section{Simulation Setup}\label{sec:setup}
We use the Message Passing Interface - Adaptive Mesh Refinement Versatile Advection Code (MPI-AMRVAC) to solve the MHD equations on an axisymmetric spherical (2.5 \,D) domain to simulate a breakout CME. The domain ranges from 1 $R_\odot$ to 10 $R_\odot$ in the radial direction and from the north to the south pole in the angular direction, i.e., (r,$\theta$) $\epsilon$ $[1 R_\odot, 10 R_\odot]$ $\times$ [0, $180^\circ$], where r is the radial distance from the center of the sun and $\theta$ is the heliographic colatitude. Initially, the domain is divided into a grid with a resolution of 128 $\times$ 128 uniformly stretched cells (the ratio between the length and the width of the cells is kept constant). This means that as one moves away from the inner boundary, the cell size increases at a constant scale. Further, in order to achieve a high resolution to see the dynamics of the system in great detail, we use AMR by adding three more levels to the initial resolution. We choose the refinement criteria in such a way that allows us to refine the regions with strong current density. For this, we calculate a dimensionless quantity \citep{Karpen2012,Hosteaux2018}:
\begin{equation}
c \equiv \frac{\left|\iint_{S} \nabla \times \boldsymbol{B} \cdot \mathrm{d} \boldsymbol{a}\right|}{\oint_{C}|\boldsymbol{B} \cdot \mathrm{d} \boldsymbol{l}|}=\frac{\left|\oint_{C} \boldsymbol{B} \cdot \mathrm{d} \boldsymbol{l}\right|}{\oint_{C}|\boldsymbol{B} \cdot \mathrm{d} \boldsymbol{l}|}=\frac{\left|\sum_{n=1}^{4} B_{t, n} l_{n}\right|}{\sum_{n=1}^{4}\left|B_{t, n} l_{n}\right|} .
\end{equation}
where \textit{B$_{t,n}$} is the tangential component of the magnetic field \textbf{B} along the $n$th segment $l_n$ of the contour \textit{c} of a cell. The refinement parameter $c$ is the ratio of the magnitude of the electric current that passes through the surface $S$ enclosed by the contour $C$ of a cell and the sum of the absolute value of individual contributions to the current. We can see that $c$ can take a maximum value of 1. The strong current carrying regions have a value near 1. In our setup, if \textit{c} has a value less than 0.01, we coarsen the block to a lower level, and if \textit{c} is greater than 0.02, the block is refined. Between these two values, AMR routine does not enforce any constraints, allowing the grid to maintain its resolution from preceding time steps.\par
We use the total variation diminishing Lax-Friedrichs (TVDLF) scheme with minmode slope limiter. We use the generalised Lagrange multiplier \citep[GLM,][]{Dedner2002} method to keep the magnetic field divergence free, which transports any unphysical magnetic monopoles outside of the domain. The Courant–Friedrichs–Lewy (CFL) number was kept 0.3 while solving the MHD equations. Units are scaled with a scaling factor which, for length is the solar radius, for velocity is 100 km s$^{-1}$, and for time is 1.93 h.\par
Our simulation setup includes a background global dipole magnetic field, a multipolar arcade system and shearing velocity for helicity injection, which is applied only when the system reaches the equilibrium. The gravitational field was obtained by introducing an extra source term to the momentum equation. The fast and slow bimodal distribution of the solar wind \citep{Jacobs2005, chene2006, chene2008, Hosteaux2019} is reproduced by the heating term added to the MHD equations. This empirical heating source term is a function of latitude and radial distance from the center of the Sun. This volumetric heating term is of the following form{\citep{Groth2000,manchester2004} :
\begin{equation}
    Q = \rho q_0(T_0-T)\exp\biggl[\frac{-(r-1 R_\odot)^2}{\sigma_0^2}\biggr]
\end{equation}
where $q_0$ = 10$^6$ ergs g$^{-1} $s$^{-1} $K$^{-1}$ is the volumetric heating amplitude, $\rho$ is the mass density, $R_\odot$ is the solar radius, and $T(K)$ is the plasma temperature. Target temperature $T_0(K)$ and heating scale height $\sigma_0$ ($R_\odot$) depend on the value of the critical angle $\theta_0$, which is measured from the north pole. $T_0$ takes the value of 1.5 $\times$ 10$^6$ K equator-ward and 2.63 $\times$ 10$^6$ K pole-ward from this critical angle $\theta_0$. The value of the heating scale height $\sigma_0$ is 4.5 R$_\odot$ towards the equator and 4.5[2 - sin$^2\theta$/sin$^2\theta_0$] R$_\odot$ pole-ward from this critical angle $\theta_0$. The critical angle is defined as sin$^2\theta_0$ = sin$^2(17.5^\circ$) + cos$^2(17.5^\circ$)(r/R$_\odot$ - 1)/8 for r $\leq$ 7 R$_\odot$, while for r $>$ 7 R$_\odot$ this value changes to sin$^2\theta_0$ = sin$^2(61.5^\circ$) + cos$^2(61.5^\circ$)(r/R$_\odot$ - 7)/40.\par
For the boundary conditions and the initial conditions, we fix the number density to 10$^8$ cm$^{-3}$ and temperature to 1.5 $\times$ 10$^6$ K at the inner boundary. To include the bimodal distribution of background solar wind, we start with a constant radial component of the velocity $v_r$ but with the model used in \citep{Jacobs2005, chene2006,chene2008,Hosteaux2019} and with the modification in the MHD equation by using the heating term we get the fast and the slow solar wind profile. This radial component of velocity $v_r$ is set to zero in the ghost cells. $v_\theta$ is set to zero at the inner boundary. We impose the differential rotation in the form of longitudinal velocity component $v_\phi$, which is set accordingly at the inner boundary. To get the background bipolar magnetic field profile, we fix the $r^2B_r$ at the inner boundary. Moreover, we extrapolate the $r^5B_\theta$ \citep[introduced by][]{dana2020,dana2022_propagation} and $B_\phi$. All the variables $r^2\rho$, $r^2\rho v_r$, $\rho v_{\theta}$, $rv_\phi$, $r^2B_r$, $B_\theta$, $rB_\phi$, and $T$ at the outer boundary are kept continuous.\par

\begin{figure*}
\gridline{\fig{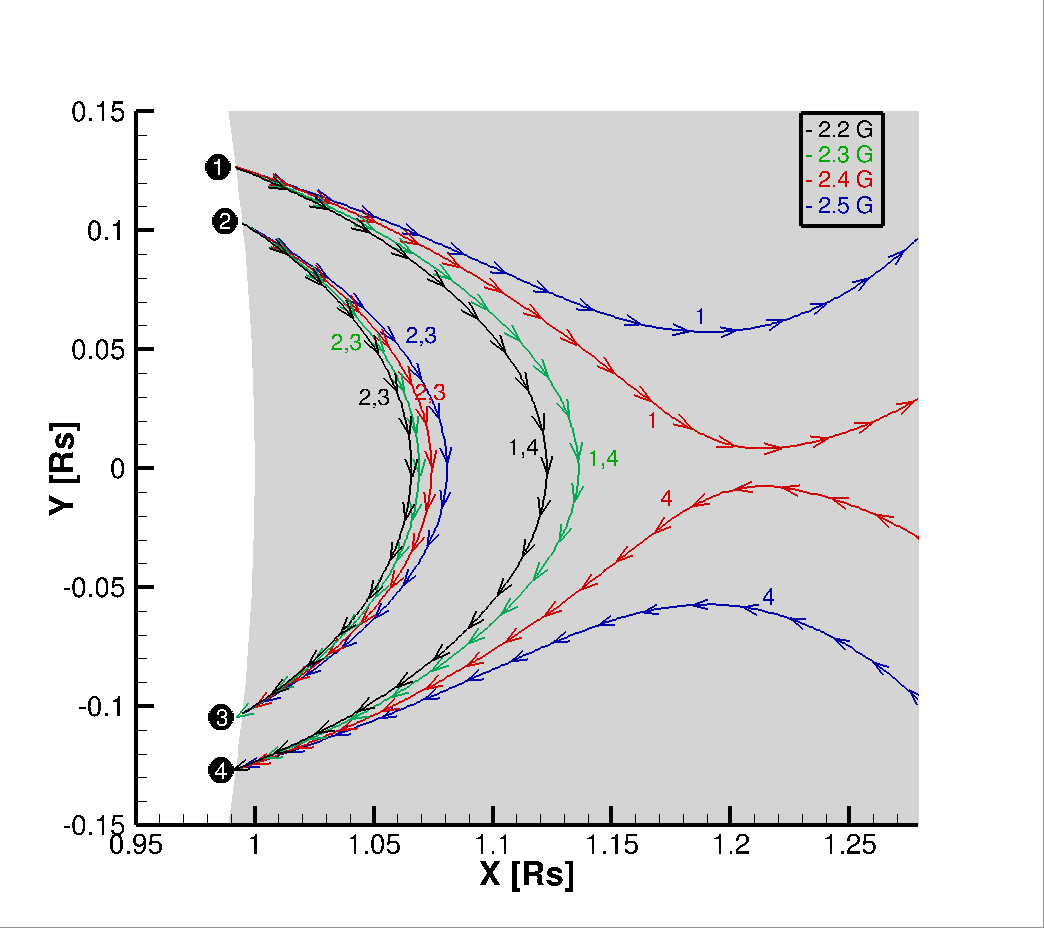}{0.45\textwidth}{(a)}
          \fig{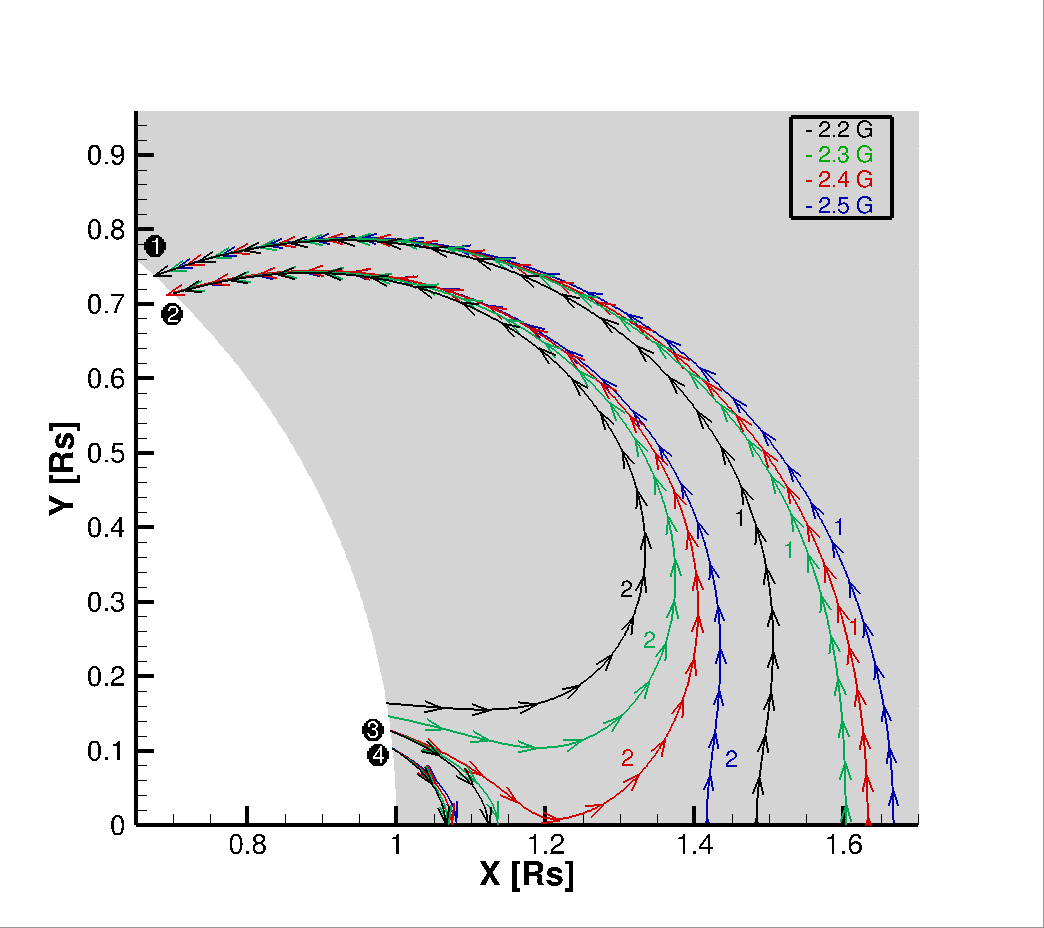}{0.45\textwidth}{(b)}}
\caption{Selected magnetic field lines
tracing in the central and northern arcade for different values of background poloidal magnetic field at the poles at the steady state. \label{equil}}
\end{figure*}

The background dipole magnetic field components has the following form:
\begin{equation}\label{br}
    B\textit{dip}_{r}= \frac{2B\cos\theta}{r^3}
\end{equation}
\begin{equation}
    B\textit{dip}_{\theta}= \frac{B\sin\theta}{r^3}
\end{equation}
Here, $\textit{B}$ is a constant that determines the strength of the background dipole magnetic field. We added a triple arcade system to create a quadrupole magnetic field topology similar to those reported in \citet{van_der_Holst_2007} and \citet{Zuccarello_2012}. The vector potential for this system is taken from \cite{van_der_Holst_2007} and has the following form:

\begin{equation}
    A_\phi = A_0\frac{\text{cos}^2(180^{\circ}\lambda/2\Delta a)}{r^4\text{sin}\theta}
\end{equation}
By taking the curl of this vector potential, we obtained the following magnetic field components
\begin{equation}
    B_r = \frac{A_0}{2r^5\sin\theta}\frac{180^{\circ}}{\Delta a}\text{sin}\biggl[\frac{180^{\circ}\lambda}{\Delta a}\biggr],
\end{equation}
\begin{equation}
    B_\theta = \frac{3A_0}{r^5\sin\theta}\text{cos}^2\biggl[\frac{180^{\circ}\lambda}{2\Delta a}\biggr],
\end{equation}

\noindent where the solar latitude is represented by $\lambda$ = 90$^\circ$ - $\theta$, $A_0$ = -0.73 G $R_\odot^5$ and $\Delta a$ = 28.64$^{\circ}$. This additional term is only applied to the dipole field when the absolute value of $\lambda$ is less than 28.64$^{\circ}$ ($\lvert \lambda \rvert$ $<$ 28.64$^{\circ}$). We first allow the system to reach a steady state. After the system reaches this steady state, we then impose time-dependent shearing at the inner boundary, which takes the following form:
\begin{equation}
    v_\phi = v_0(\lambda^2 - \Delta b ^2)^2\text{sin}\lambda~\text{sin}[180^\circ(t - t_0)/\Delta t]
\end{equation}
We imposed the shearing velocity in the latitude range where $|\lambda|$ $<$ 8.59$^{\circ}$, which is located well within the central arcade. Anywhere outside this range, the $v_\phi$ is set to zero. We kept $\Delta b$ = 8.59$^{\circ}$. The running time is denoted by $t$. The duration of the shearing ($\Delta t$) was 27 hours. The time required to reach the steady state, $t_0$, is $\approx$ 200 hours in this scenario. After reaching the steady state, we reset the time counter to zero and now $t_0$ = 0 hours. Shearing velocity exhibits a gradual increase and decrease, reaching its peak at half-time of the total duration of shear. It is worth noting that the inner boundary of our setup represents the lower corona, not the photosphere. Also, $v_0$ is chosen such that maximum $v_\phi$ at $t$ = 13.5 h does not exceed 10\% of the local Alfv$\acute{\text{e}}$n speed. The local Alfv$\acute{\text{e}}$n speed within the central arcade near the inner boundary ranges from 600 to 750 km s$^{-1}$. The maximum shearing velocity required to trigger an eruption is also linked to the grid resolution, as magnetic reconnection becomes effective only when current sheets reach scales comparable to the local grid size. Consequently, increasing the resolution lowers the maximum shearing velocity needed to initiate an eruption.

\section{Numerically simulated eruptions}
We applied the shear motion at the base of the middle arcade in the region where the latitude range is $|\lambda|$~$<$ 8.59$^{\circ}$. The shearing motion increases $v_\phi$, and therefore also the $\phi$ component of the magnetic field B$_\phi$ resulting in the disparity between the magnetic pressure and magnetic tension. This imbalance leads to the expansion of the central arcade, which flattens the X-point above it, allowing magnetic reconnection to take place due to numerical resistivity. This reconnection process, known as the breakout reconnection, results in the disconnection and ejection of the top of the helmet streamer.

\subsection{Effect of the poloidal magnetic field strength}
We explore the influence of altering the strength of the poloidal magnetic field by varying the background poloidal magnetic field strength at the poles from 2.2 G to 2.5 G with a step size of 0.1 G. Figure \ref{equil} illustrate the magnetic field configurations of the central and northern arcade systems for various values of the background poloidal magnetic field at steady state. In Figure \ref{equil}(a), magnetic field lines originating from footpoint 1 exhibit different configurations depending on the background poloidal magnetic field strength at the pole. For values of 2.2 G and 2.3 G, these field lines form part of the central arcade system. However, for higher field strengths of 2.4 G and 2.5 G, the field lines reconnect and become part of the northern arcade system, highlighting the influence of the background magnetic field. The magnetic field lines originating from footpoint 2 consistently remain within the central arcade system across all values of the background poloidal magnetic field. Similarly, in Figure \ref{equil}(b), the magnetic field lines ending at the same footpoints come from distinct paths based on the background poloidal magnetic field strength. Each colour-coded magnetic field line in the figure corresponds to a specific value of the initial background poloidal magnetic field strength at the poles.}
For a given poloidal magnetic field of 2.2 G at the poles, we observe an elevation in the central arcade of the multipolar arcade system, leading to a breakout CME (Figure \ref{pol_field_images_second}(a)). With a poloidal magnetic field strength of 2.3 G at the poles, we witness the ascent of the central arcade, yet the formation of the flux rope is not apparent (Figure \ref{pol_field_images_second}(b)). This increase in the poloidal magnetic field strength not only restrains the eruption but also hinders the formation of the flux rope. 
Upon further increasing the poloidal field strength to 2.5 G at the poles, we observe a reduction in the maximum height achieved by the central arcade during the shearing motions at the base (Figure \ref{pol_field_images_second}(c), \ref{pol_field_images_second}(d)). Apart from polar magnetic field strength, all the other parameters, such as domain size and shear velocity at the base, were kept the same for all the different cases. Therefore, it is evident that the background magnetic field strength significantly influences the eruption of CMEs.

\subsection{Effect of change in the helicity injection}
\begin{figure*}[!ht]
\gridline{\fig{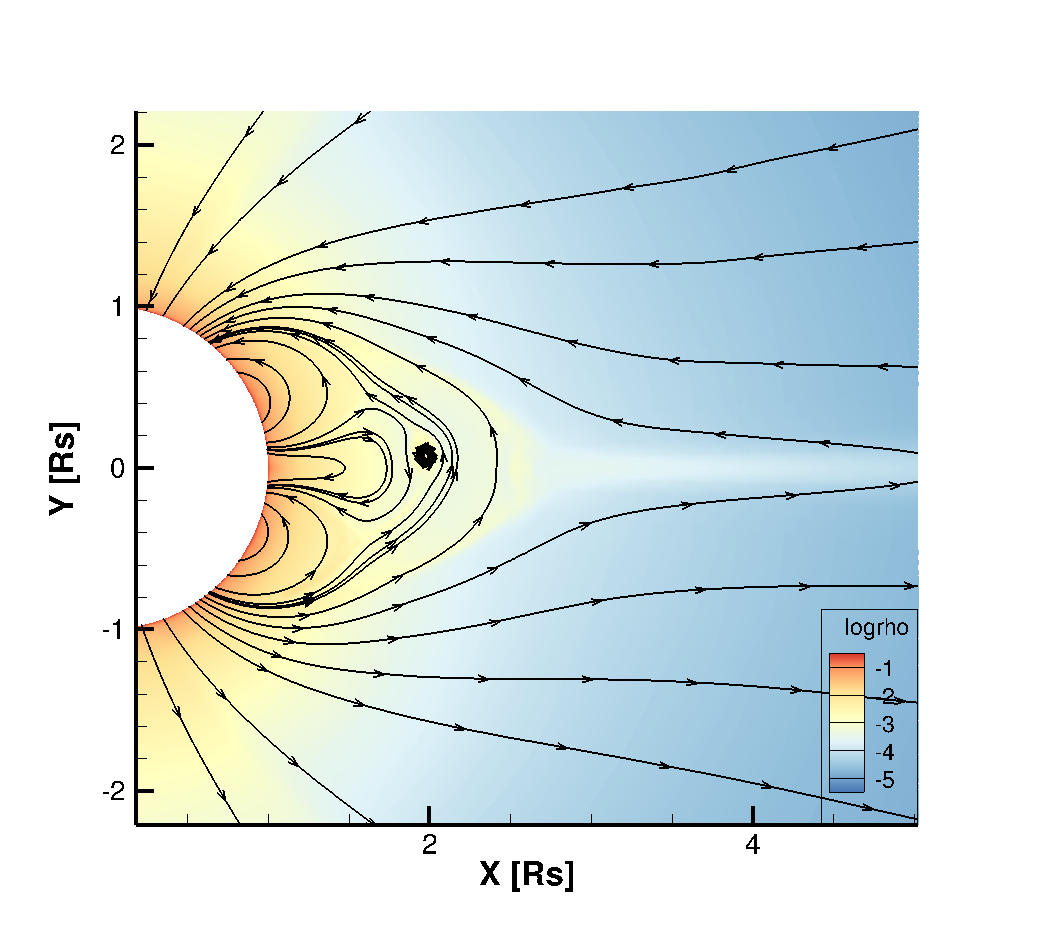}{0.48\textwidth}{(a)\label{fig:sub1_failed}}
          \fig{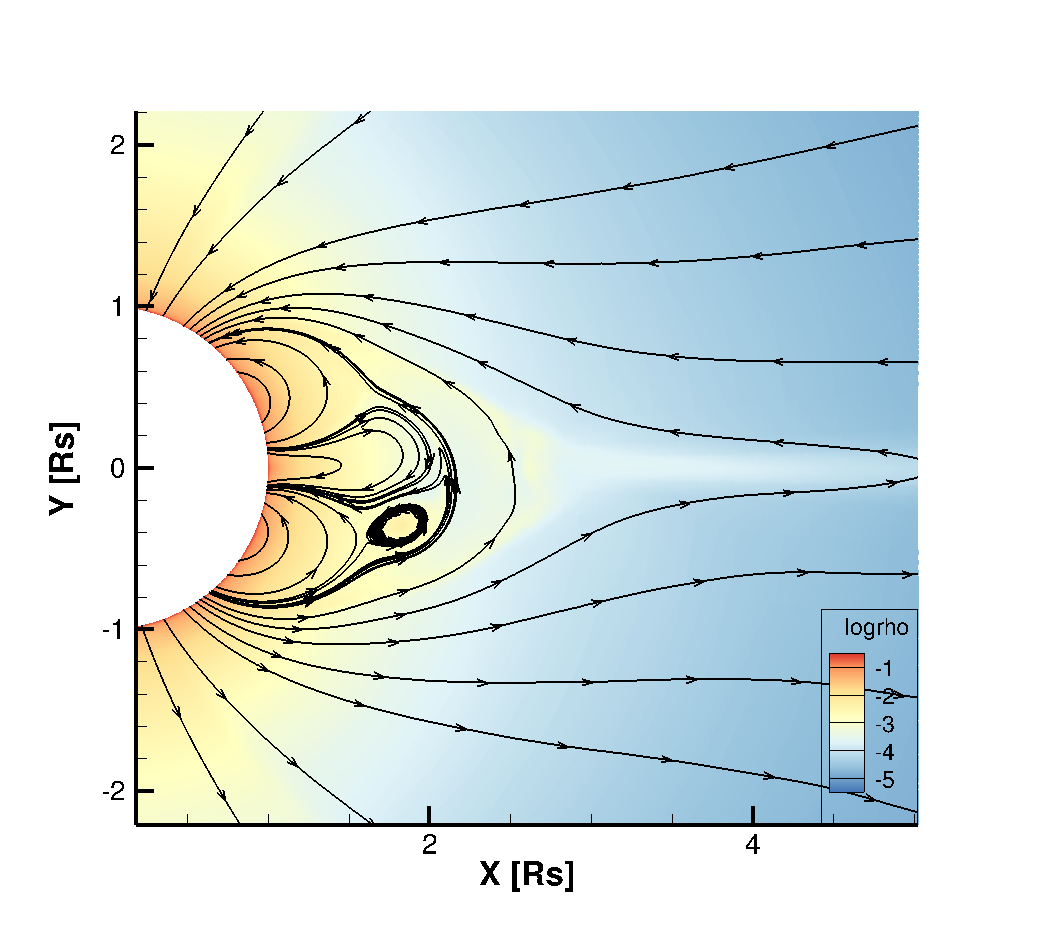}{0.48\textwidth}{(b)\label{fig:sub2_failed}}}

\caption{Simulation snapshots depicting selected magnetic field line traces (black lines) and density in log scale (colour scale). These panels represent the failed eruption case, and at the following times from the start of shear: (a) at $t$ = 14.68 h (b) at $t$ = 15.46 h. A movie of temporal evolution of the corresponding density is available as supplementary material 1.\label{fig:failed_eruption}}   
\end{figure*}

\begin{figure*}[!ht]
\gridline{\fig{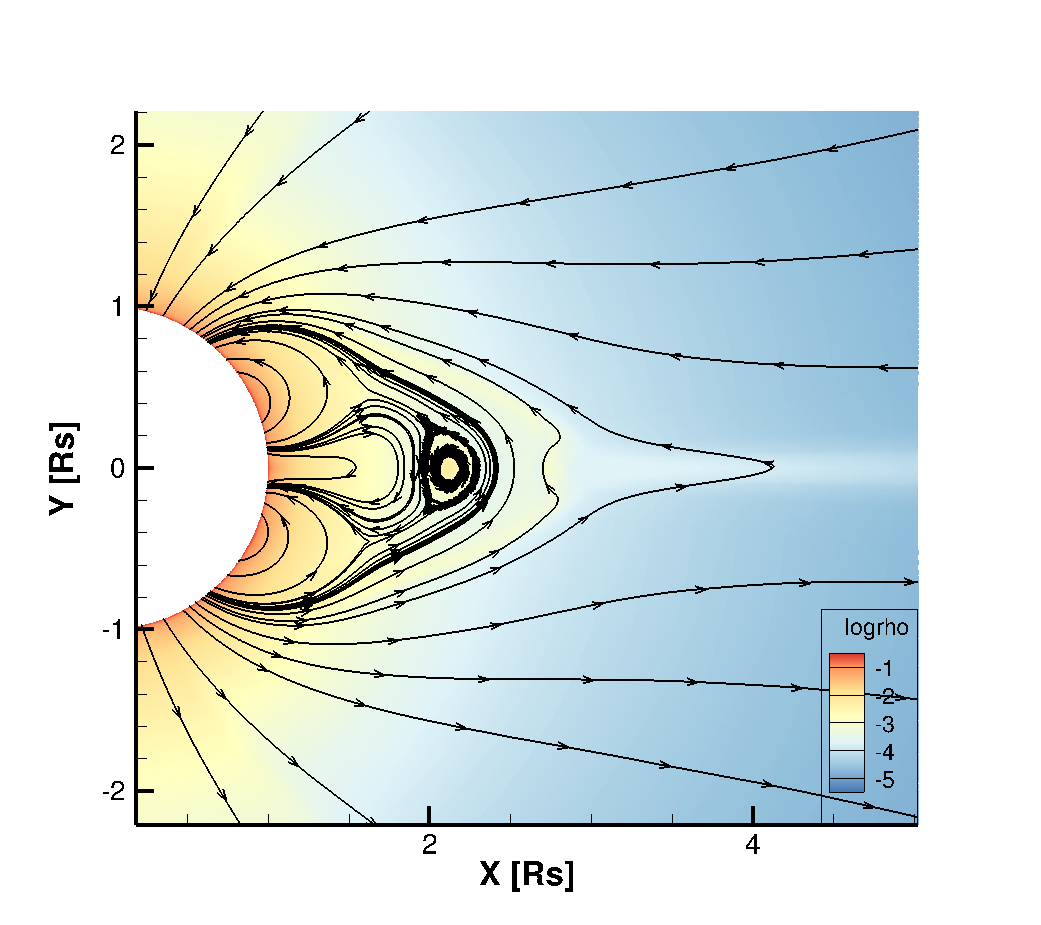}{0.48\textwidth}{(a)\label{fig:sub1_single}}
          \fig{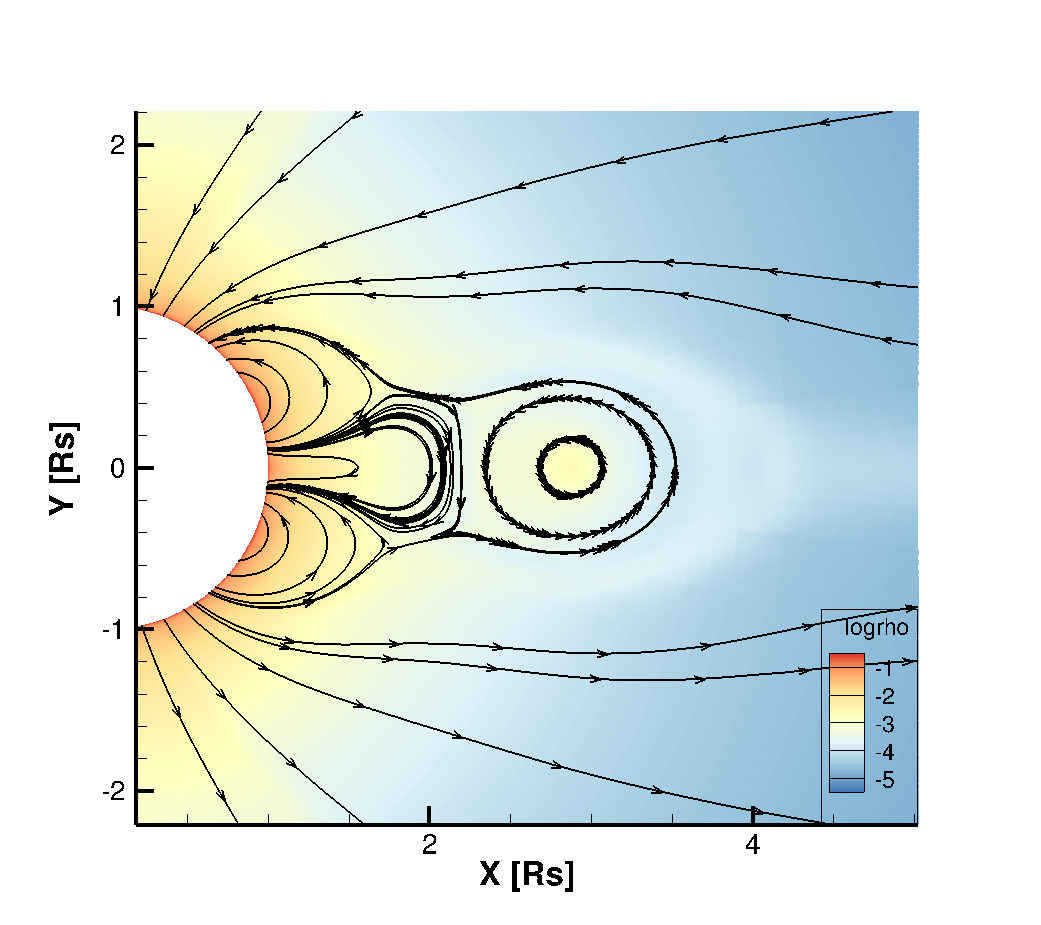}{0.48\textwidth}{(b)\label{fig:sub2_single}}}

\caption{Same as Fig. \ref{fig:failed_eruption}, but for the single eruption case, and at the following times from the start of shear: (a) at $t$ = 12.368 h (b) at $t$ = 16.233 h. A movie of temporal evolution of the corresponding density is available as supplementary material 2. \label{fig:single_eruption}}
\end{figure*}

\begin{figure*}
\gridline{\fig{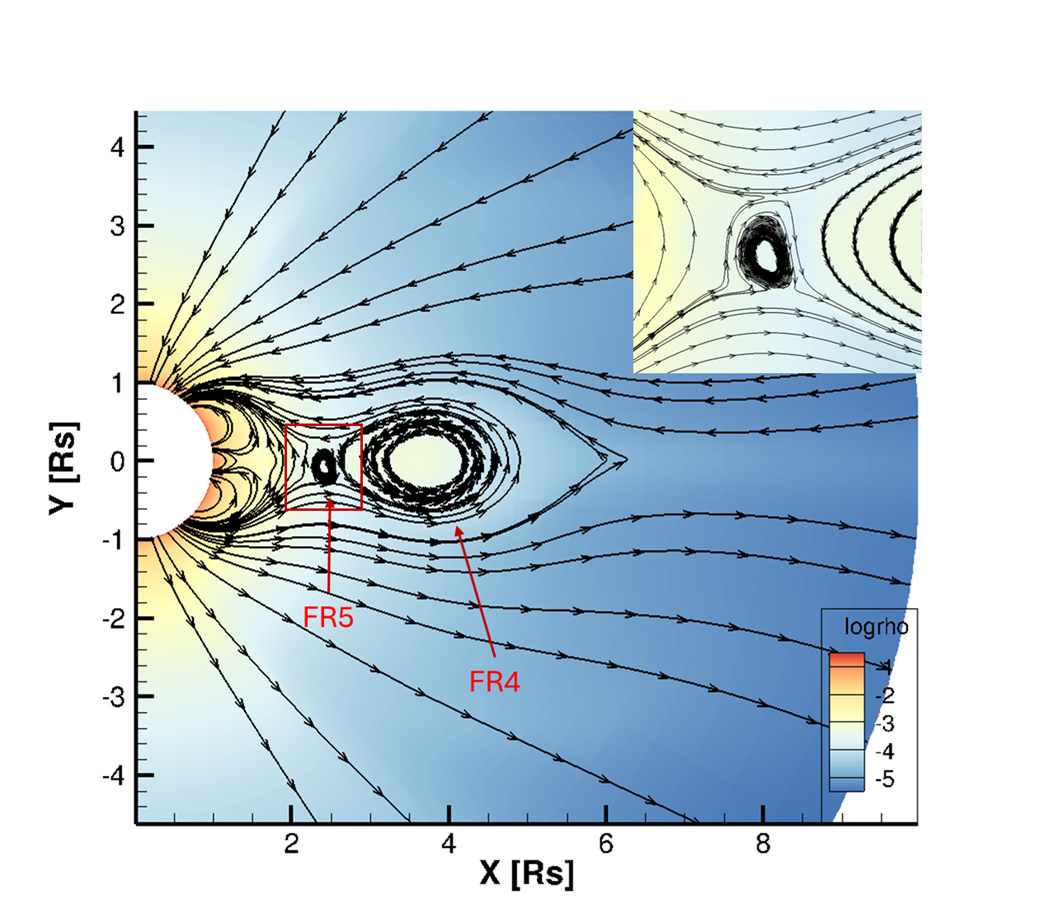}{0.45\textwidth}{(g) t = 26.48 h\label{fig:sub7_multi}}
          \fig{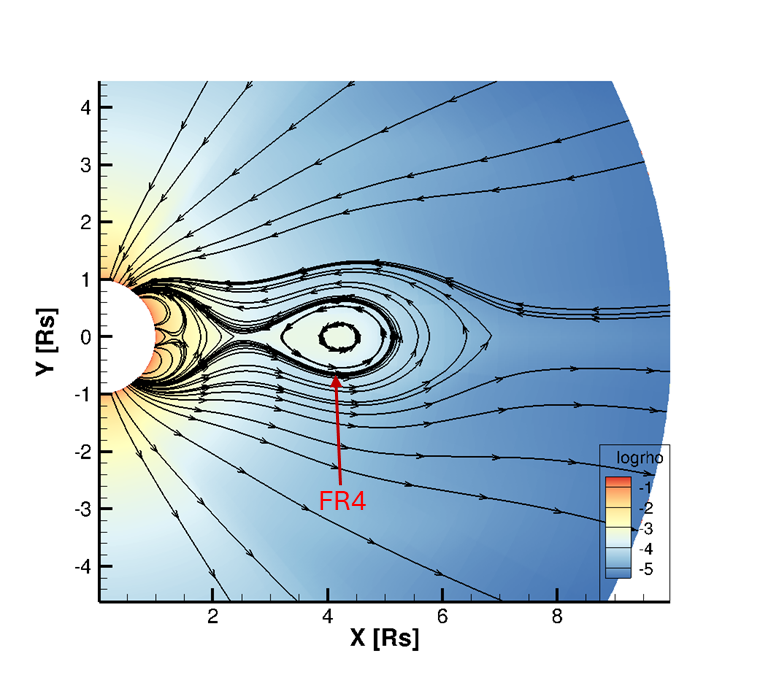}{0.45\textwidth}{(h) t = 28.02 h\label{fig:sub8_multi}}}
\caption{Same as Figs. \ref{fig:failed_eruption} and \ref{fig:single_eruption}, but for the multiple eruptions case. In (b), (c), (f), and (g), the top right corner displays a zoomed-in view of the area marked with a red box. A movie of temporal evolution of the corresponding density is available as supplementary material 3. \label{fig:multi_eruption}}
\end{figure*}

The magnitude of shear velocity at the base of the magnetic field arcade system governs the injection of helicity into the arcade system. For a fixed value of 2.2 G of the radial background magnetic field at the pole ($B_{pol}$), the three distinct values of maximum shear velocity at the base lead to three different scenarios for arcade evolution.
\begin{enumerate}[itemsep=1pt]
\item The failed eruption: For a shear velocity of 35.8 km s$^{-1}$ at the base, we observe a failed eruption. Although a flux rope is generated as a result of the shearing motion at the base (Figure \ref{fig:failed_eruption}(a)), it subsequently descends back to the surface of the Sun (Figure \ref{fig:failed_eruption}(b)) along the magnetic field lines towards the southern arcade. The temporal evolution of the corresponding density can be seen as supplementary material 1. 
\item The single eruption: Increasing the maximum shearing velocity at the base to 36.2 km s$^{-1}$ we obtain an eruption. The flux rope forms as a result of magnetic reconnection (Figure \ref{fig:single_eruption}(a)) and the breakout reconnection facilitates its eruption (Figure \ref{fig:single_eruption}(b)). The temporal evolution of the corresponding density can be seen as supplementary material 2.

\begin{figure*}[!ht]
\gridline{\fig{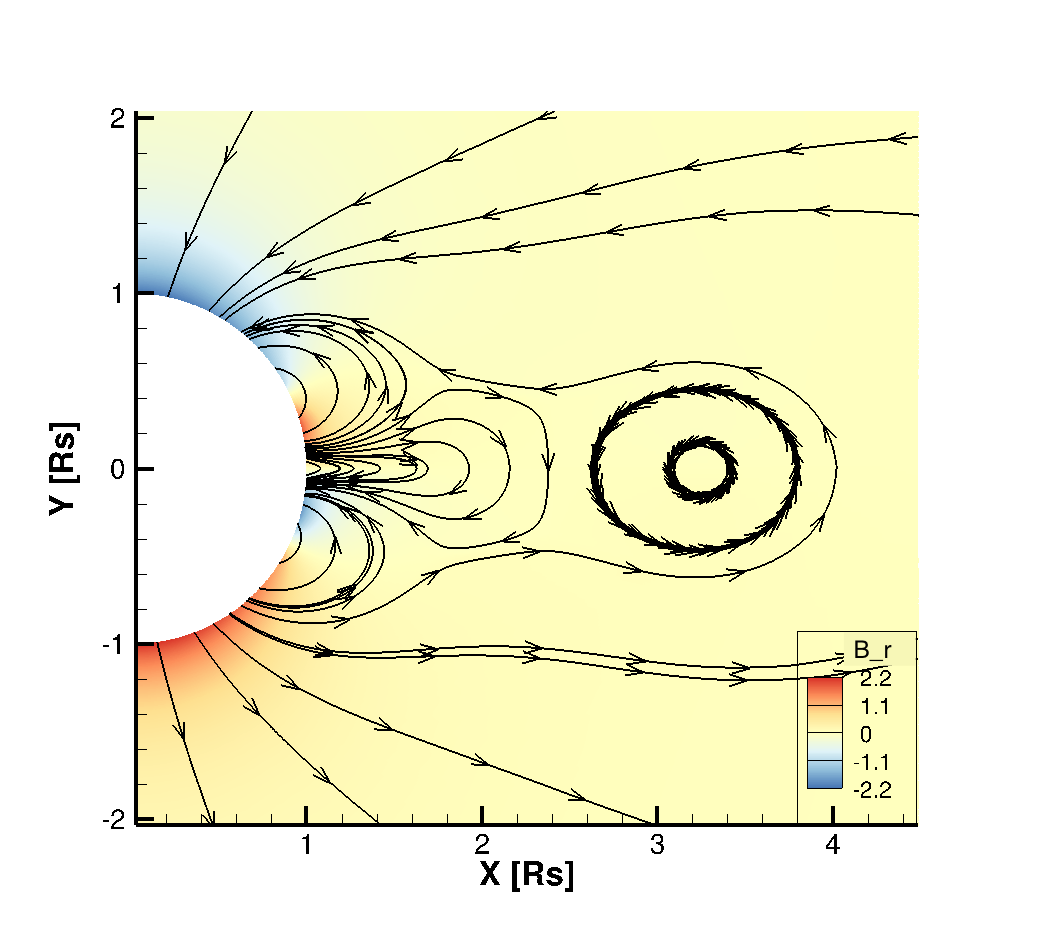}{0.45\textwidth}{(a) t = 15.46 h; B$_{pol}$ = 2.2 G\label{fig:sub1}} 
          \fig{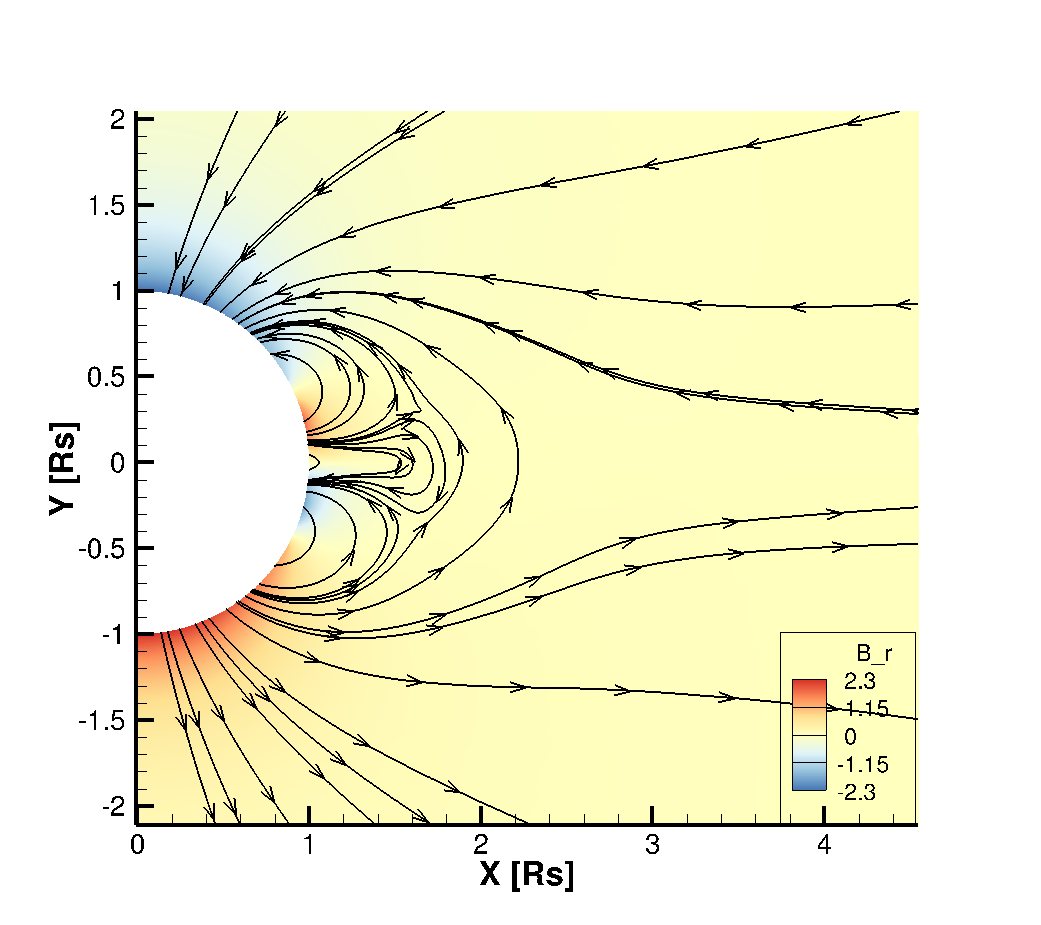}{0.45\textwidth}{(b) t = 11.59 h; B$_{pol}$ = 2.3 G\label{fig:sub2}}}
\gridline{\fig{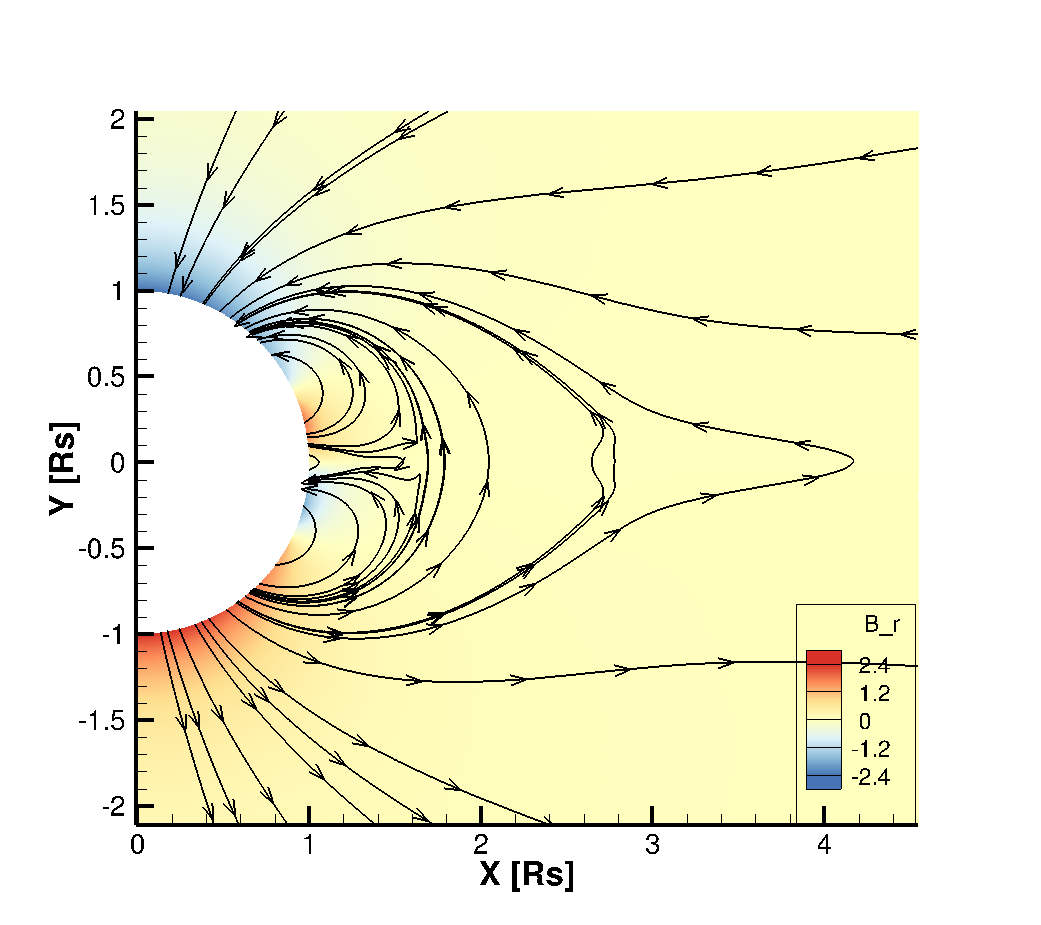}{0.45\textwidth}{(c) t = 10.04 h; B$_{pol}$ = 2.4 G\label{fig:sub3}}
          \fig{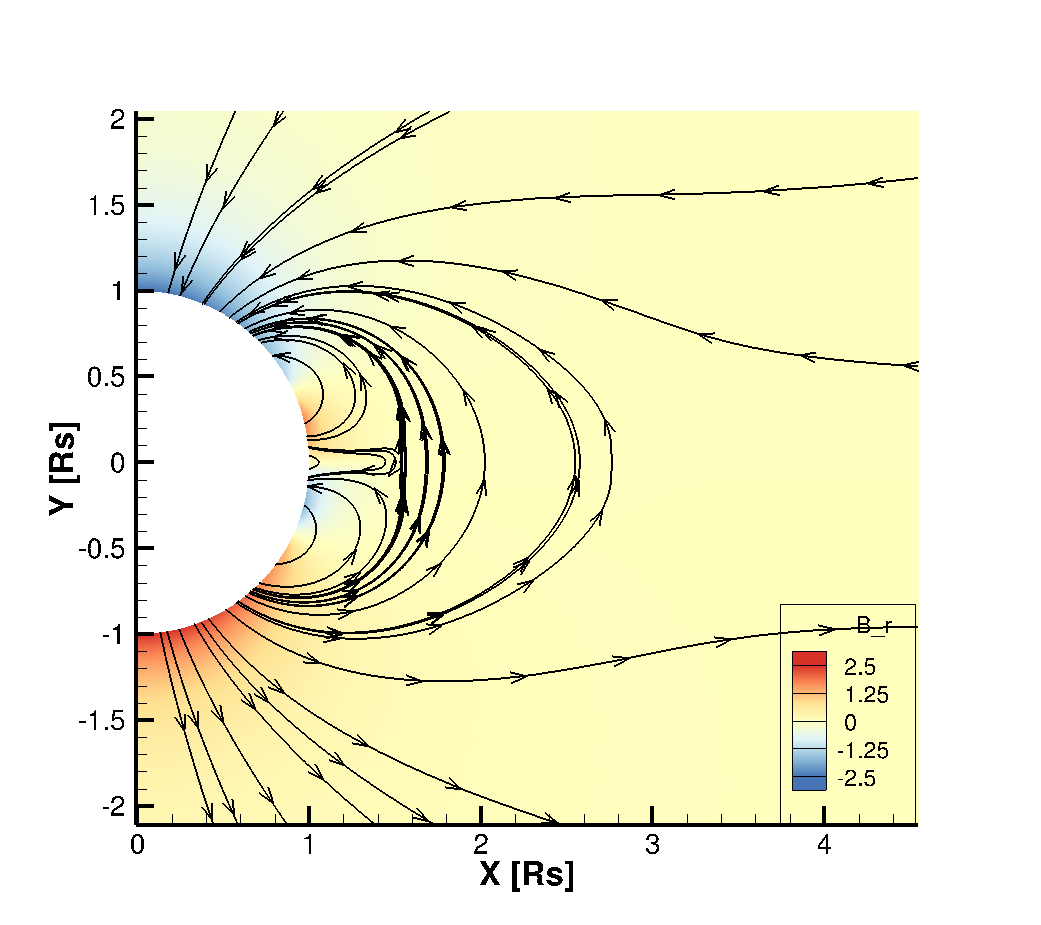}{0.45\textwidth}{(d) t = 8.50 h; B$_{pol}$ = 2.5 G\label{fig:sub4}}}

\caption{Selected magnetic field lines
tracing and simulated radial component of magnetic field maps (G) at different times (t) after shearing starts, for different polar magnetic field strengths (B$_{pol}$) \label{pol_field_images_second}}
\end{figure*}
\item  The multiple eruptions:  For 38.8 km s$^{-1}$, we witness multiple eruptions. Figure \ref{fig:multi_eruption}(a) represents the first CME and Figure \ref{fig:multi_eruption}(e) represents the second CME. The flux rope eruption in both CMEs is due to the breakout reconnection process. Between these two breakout eruptions, standard flare-type reconnection also occurs, as described in the well-known CSHKP model \citep{carmichael1964,sturrock1966,hirayama1974,kopp1976,Lin2000}, leading to the formation of two other flux ropes (Figure \ref{fig:multi_eruption}(b)) and Figure \ref{fig:multi_eruption}(c). The temporal evolution of the corresponding density can be seen as supplementary material 3.
\end{enumerate}
The fundamental physical process driving breakout reconnection (indirectly) is the shearing motion occurring at the base of the magnetic arcade system. Observationally, these shearing motions exhibit speeds $\le$ 3 km s$^{-1}$ at the photosphere \citep{Grigor2007, manchester2007}. However, it's noteworthy that our domain's inner boundary starts from the base of the solar corona. Dopplergrams in H$_\alpha$ , C IV, and observations from the Solar Ultraviolet Measurements of Emitted Radiation (SUMER) instrument onboard the Solar and Heliospheric Observatory (SOHO) indicate that these shearing speeds increase with height in the solar atmosphere \citep{athay1982,malherbe,athat1985,Chae2000,manchester2007}. Specifically, they escalate from $\sim$ 5 km s$^{-1}$ in the chromosphere to 20 km s$^{-1}$ in the transition region and further to a range of 20-50 km s$^{-1}$ in the lower corona. Hence, the magnitude of the imposed shear flow matches with the velocities calculated from observations. 
\\Following the initiation of shear velocity at the base, there is an increase in the $\phi$  component of the magnetic field. This, in turn, results in an increase of the magnetic pressure within the central arcade system, leading to its expansion. This expansion pushes the neighbouring arcade, which reconnects with the overlying magnetic field, ultimately creating a passage for the flux rope. However, the flux rope does not erupt for the failed eruption scenario. In contrast, for CME,  we see the flux rope propagating outwards. With a shear velocity of 35.8 km s$^{-1}$ (Failed eruption case), it is observed that the flux rope forms approximately 14.68 hours after the initiation of shearing. Nevertheless, the flux rope does not ascend; instead, it descends back to the Sun along the magnetic field lines. Following this, the magnetic field undergoes a reconfiguration process to achieve the equilibrium. In the case of the shear velocity of 36.2 km s$^{-1}$ (Single eruption case), the flux rope forms after 12.368 hours, which is approximately 2.3 hours earlier than the failed eruption case. As previously mentioned, the flux rope formation process resembles that of the failed eruption case. However, in this scenario, the flux rope propagates outward into the current sheet in the streamer. For the maximum shear velocity of 38.8 km s$^{-1}$ (multiple eruptions case), the first flux rope (FR1) (Figure \ref{fig:multi_eruption}(a)) again forms after t = 9.66 hours, which is 2.7 hours earlier than the previously discussed single eruption case. Continued shearing causes the middle arcade to pinch off, forming another flux rope (FR2)(Figure \ref{fig:multi_eruption}(b)) at time t = 16.42 hours from the start of the shear, which inflates again and generates a subsequent flux rope (FR3) (Figure \ref{fig:multi_eruption}(c)) through magnetic reconnection, as described by the CSHKP model. These two flux ropes (FR2 and FR3) (Figure \ref{fig:multi_eruption}(c)) merge with each other and subsequently with the southern flank of the first CME due to their opposing orientations (Figure \ref{fig:multi_eruption}(d)). A similar sequence occurs for the second CME. The ongoing inflation of the middle arcade pushes against the streamer top, leading to the creation of a flux rope (FR4) (Figure \ref{fig:multi_eruption}(e)) via breakout reconnection, which then becomes the second CME. Additionally, the sheared middle arcade pinches off again, forming another flux rope (FR5) (Figure \ref{fig:multi_eruption}(f)) through CSHKP reconnection, which merges similarly with the southern flank of the second CME (Figure \ref{fig:multi_eruption}(g) and Figure \ref{fig:multi_eruption}(h)).

\hfill

\begin{figure*}[!ht]
\gridline{\fig{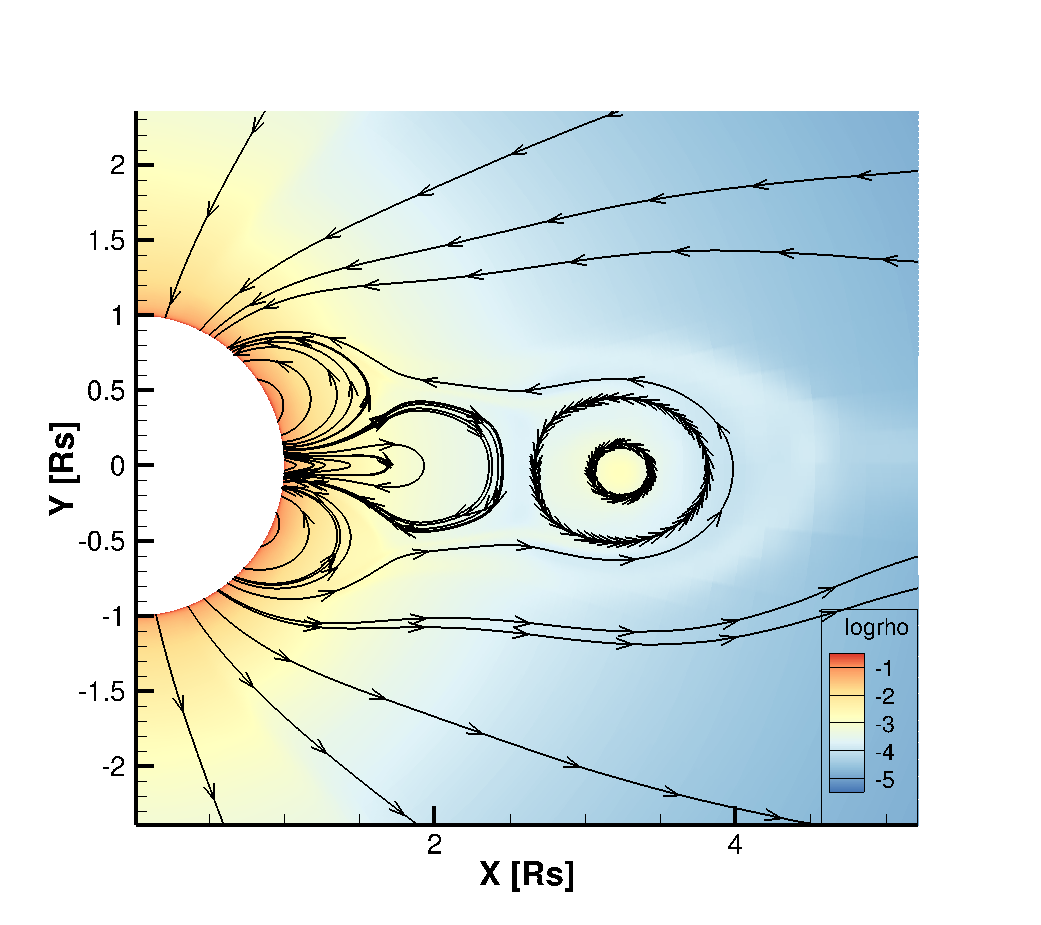}{0.45\textwidth}{(a) t = 12.56 h\label{fig:sub1_multi}}
          \fig{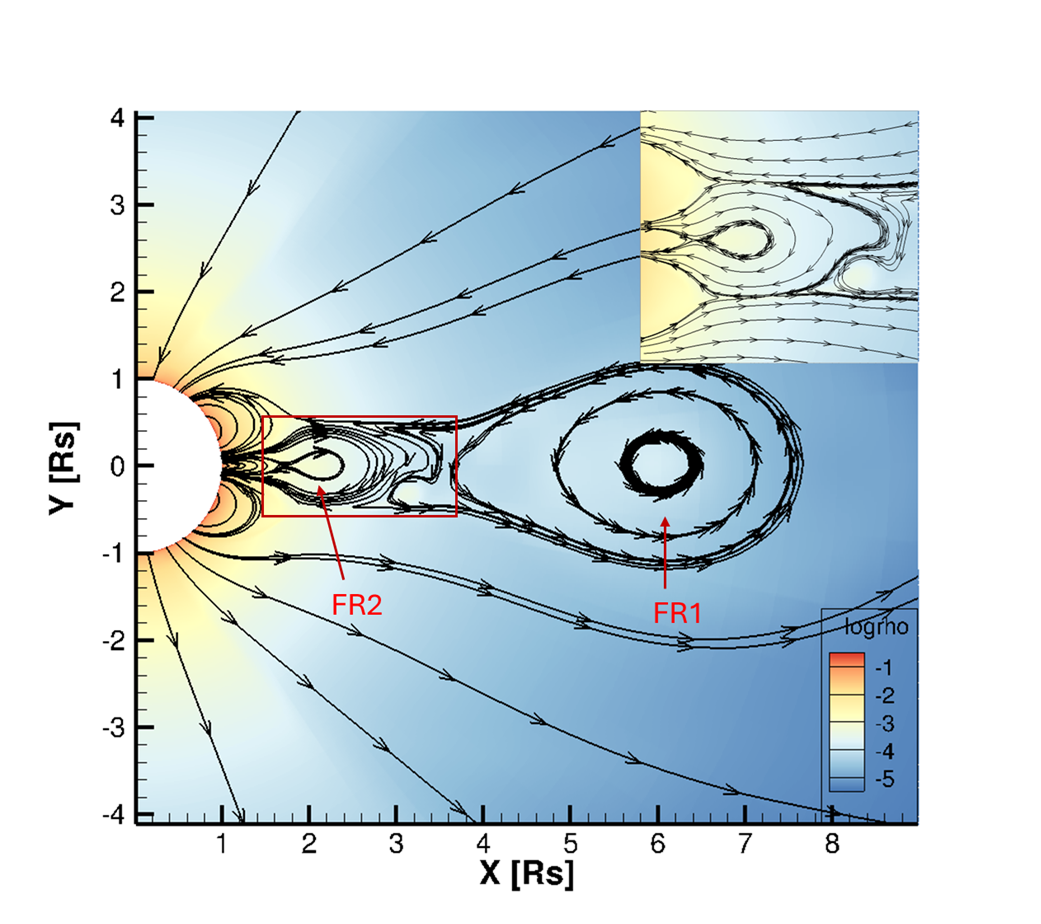}{0.45\textwidth}{(b) t = 16.42 h\label{fig:sub2_multi}}}
\gridline{\fig{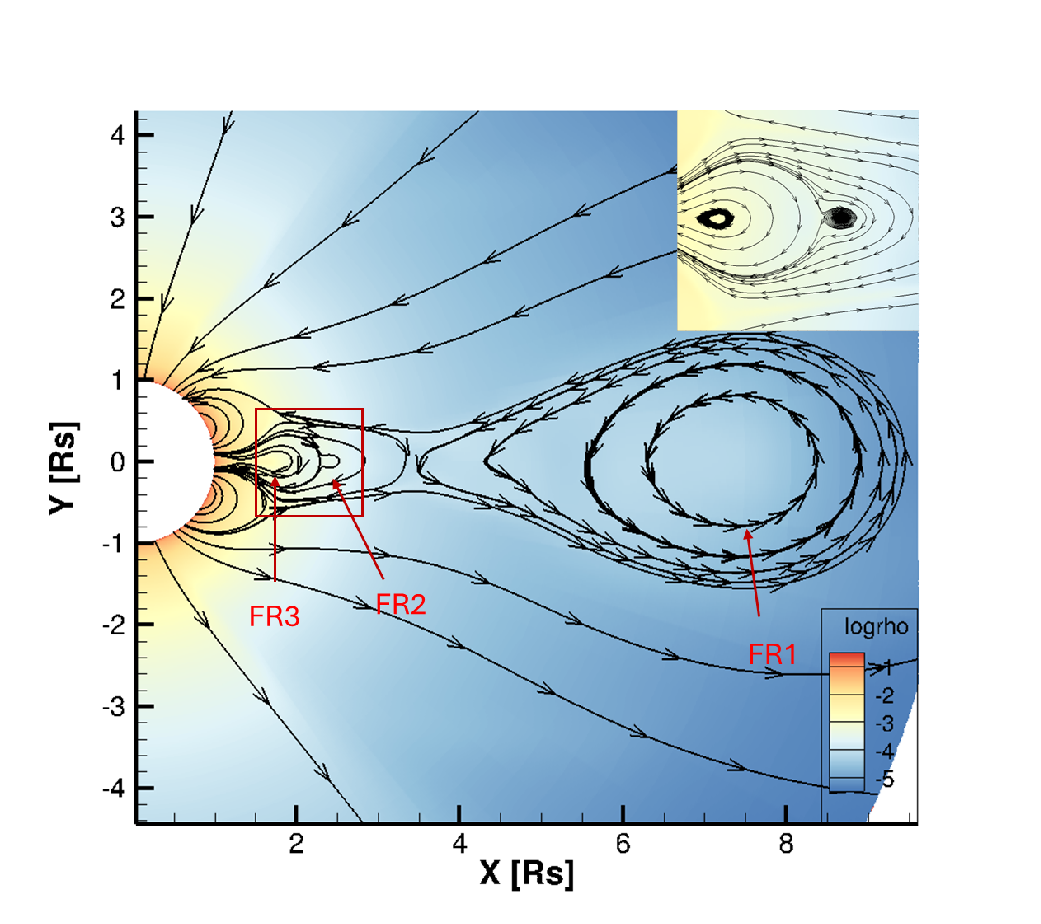}{0.45\textwidth}{(c) t = 17.58 h\label{fig:sub3_multi}}
          \fig{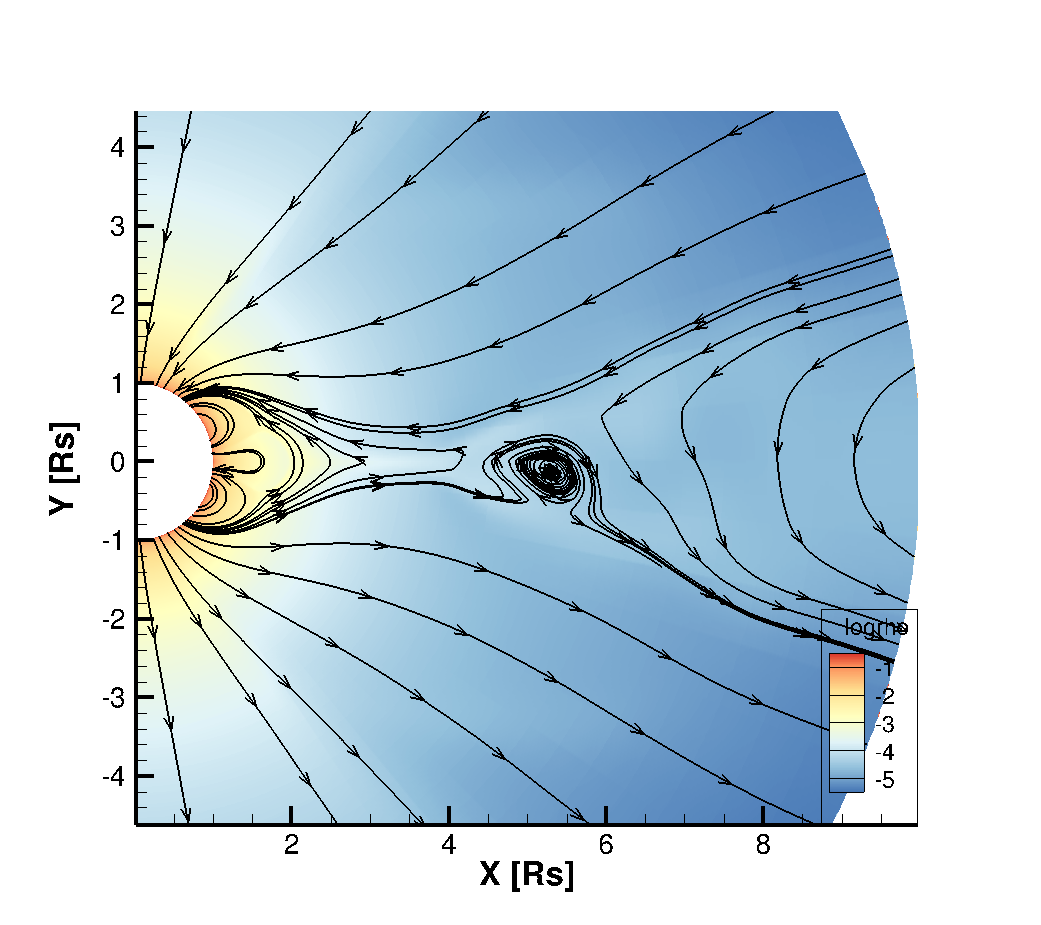}{0.45\textwidth}{(d) t = 20.29 h\label{fig:sub4_multi}}}
\gridline{\fig{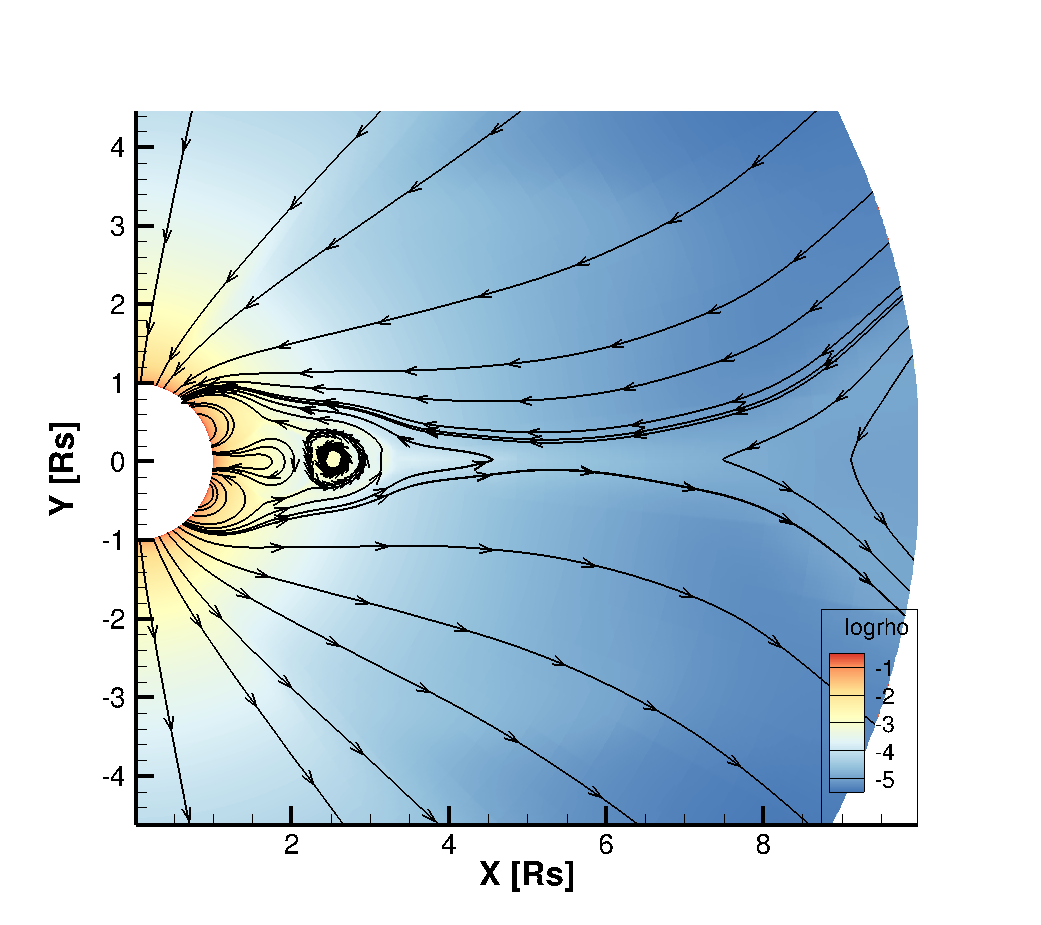}{0.45\textwidth}{(e) t = 22.22 h\label{fig:sub5_multi}}
          \fig{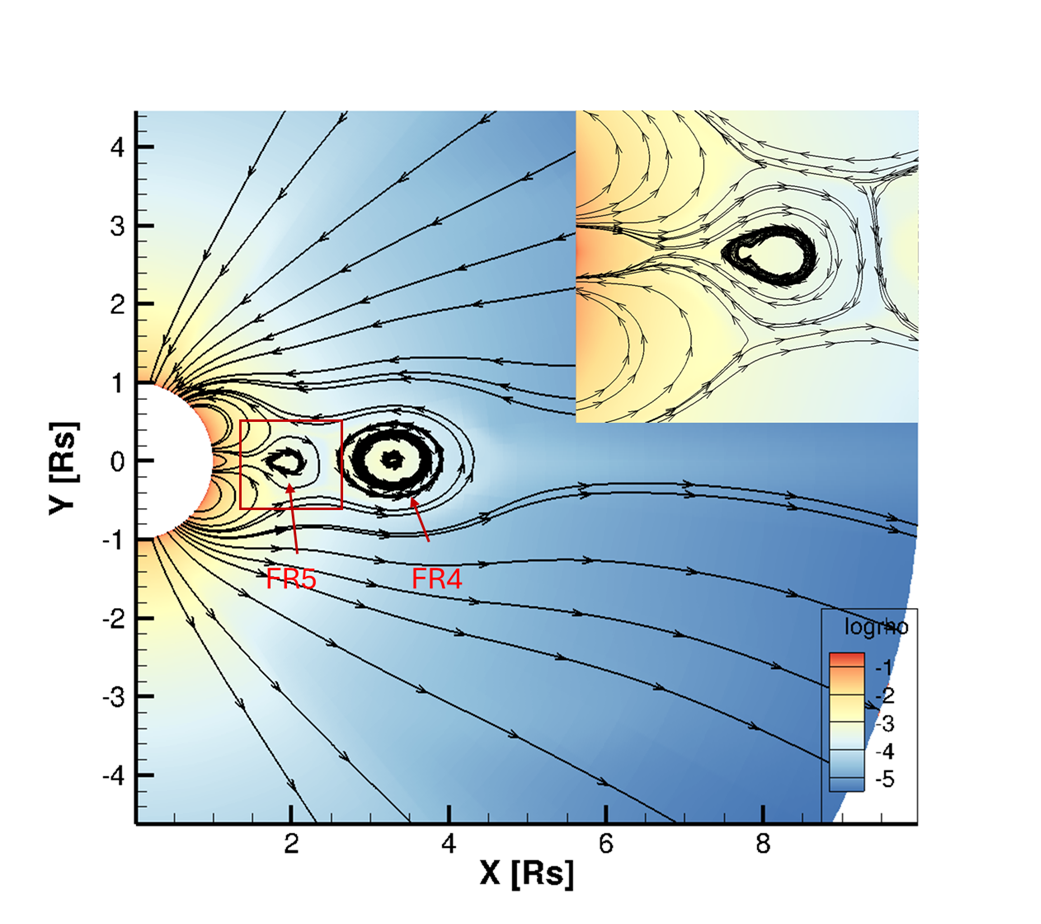}{0.45\textwidth}{(f) t = 24.92 h\label{fig:sub6_multi}}}
\end{figure*}

\begin{figure}[h!]
\gridline{\fig{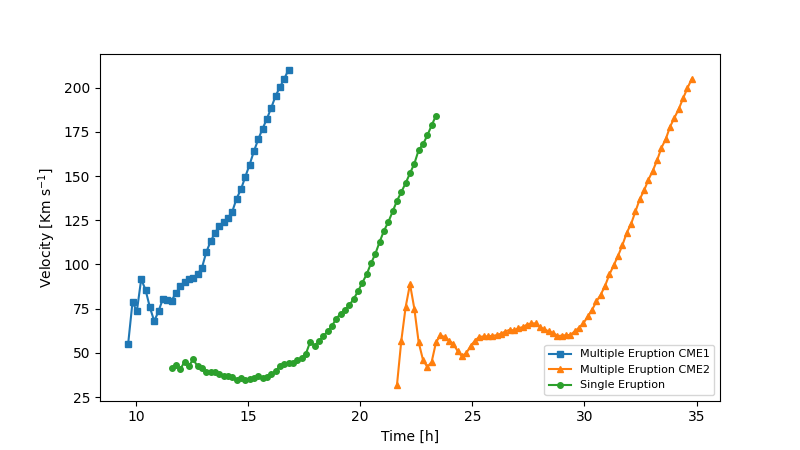}{0.5\textwidth}{ }}

\caption{Velocity-time profiles of CMEs in various eruption scenarios. The green curve illustrates the velocity of the centre of the flux rope of the  CME in the single eruption case. The blue and orange curves depict the velocity-time profiles of the centre of the flux rope associated with the first and second CMEs, respectively, in the multiple eruptions case. \label{fig:multi_velocity}}
\end{figure}

Figure \ref{fig:multi_velocity} represents the variation of the velocity of the CME core with time for different scenarios, as discussed earlier. In case of a single eruption, the velocity of the CME flux rope amounts to a total of 184.2 km s$^{-1}$, measured at a height of 6.33 R$_\odot$. In the scenario of multiple eruptions, the first CME flux rope attains a velocity of 210.1 km s$^{-1}$ at the height of 6.43 R$_\odot$. The second flux rope, formed through reconnection in accordance with the standard model, achieves a velocity of 332.4 km s$^{-1}$ at 5.97 R$_\odot$, while the third CME flux rope propagates with a velocity of 150 km s$^{-1}$ at the height of 6.31 R$_\odot$. The distances were selected such that the front of the CME does not cross the outer boundary, to avoid potential numerical influences. Notably, in the multiple eruptions scenario, the second flux rope undergoes a higher acceleration compared to the first CME. This could be attributed to the fact that while the second flux rope is forming and ascending, the magnetic field lines are stretched outwards due to the presence of the first CME. As a result, the second flux rope experiences less magnetic pressure from the overlying magnetic field. After the first CME and second flux rope pass, the magnetic field lines start to come closer, and the third CME has to propagate through the closed magnetic field lines in order to move outward. However, all CMEs formed due to breakout reconnection have a speed of less than 220 km s$^{-1}$. Thus, we can say that CMEs simulated through the shearing process are slow CMEs, within the investigated region of 10 R$_\odot$. 

\par We conduct an investigation into various global magnetic parameters across these diverse scenarios. This investigation encompasses the calculation of Total Unsigned Current Helicity (TUCH) and Absolute Net Current Helicity (ANCH) values and the total magnetic energy for each of the three different maximum shear velocity values at the base. The current helicity is defined as 

\begin{equation}
    H_c = \int_{V} \boldsymbol{B} \cdot \boldsymbol{J}\mathrm{d} \boldsymbol{V} 
\end{equation}
where $\boldsymbol{J} = (\nabla\times \boldsymbol{B})$. $\boldsymbol{B}$ is the magnetic field, and $\boldsymbol{J}$ is the current density.
The total unsigned current helicity and the absolute net current helicity is calculated as:
\begin{equation}
    TUCH =  \sum_{i=0}^{n} \lvert H_{c_i}\rvert
\end{equation}
\begin{equation}
    ANCH =  \lvert \sum_{i=0}^{n} H_{c_i}\rvert
\end{equation}
where n is the total number of cells in the region of interest.\par

\begin{figure}
\gridline{\fig{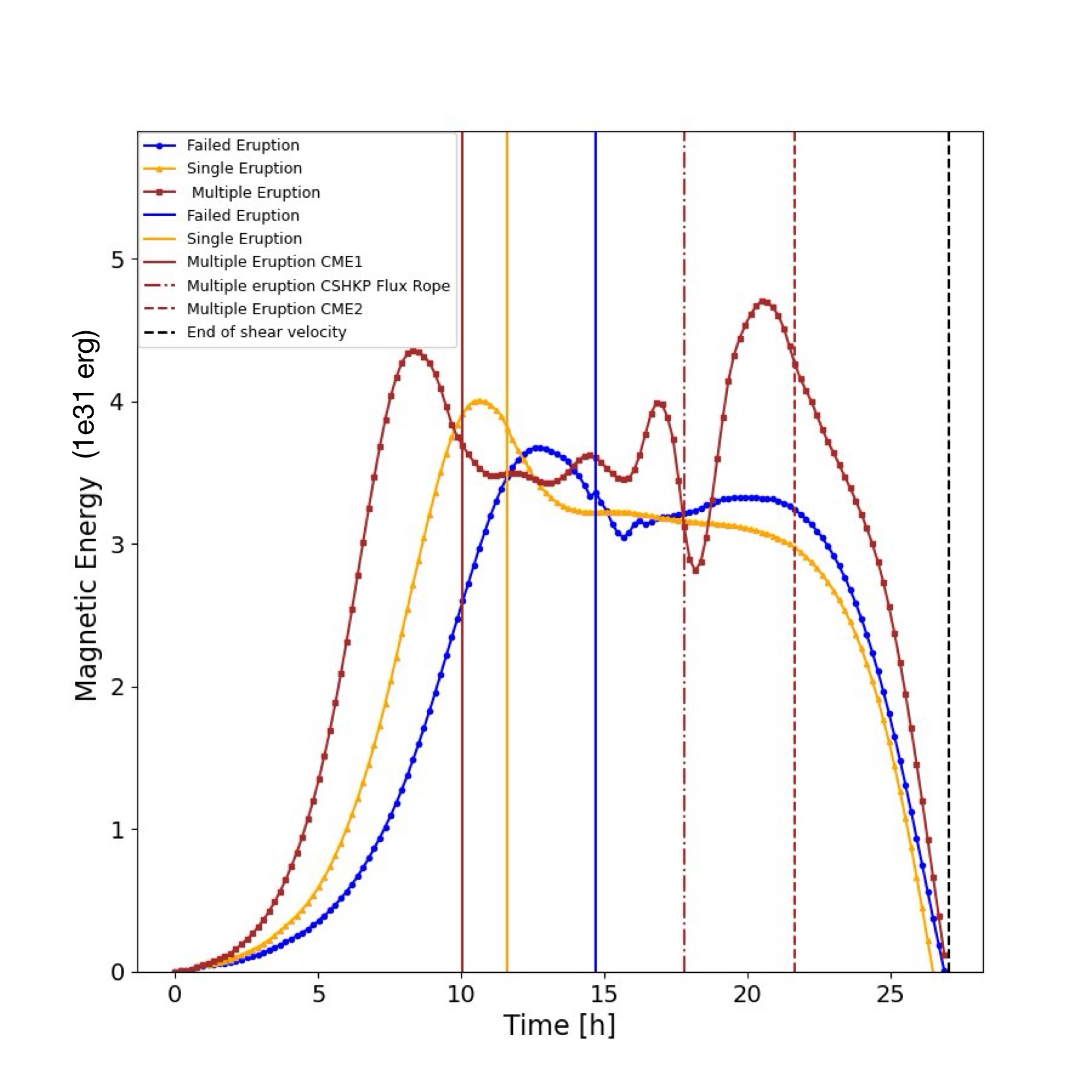}{0.42\textwidth}{ }}

\caption{Temporal evolution of excess magnetic energy for three scenarios (blue, yellow, and red for failed, single, and multiple eruptions, respectively). The blue, yellow and red vertical lines represent the flux rope formation time for the failed, single and multiple eruption cases. The black dotted vertical line represents the end of the shear. Time is measured from the start of the shear. \label{fig:me}}
\end{figure}

\begin{figure}
\gridline{\fig{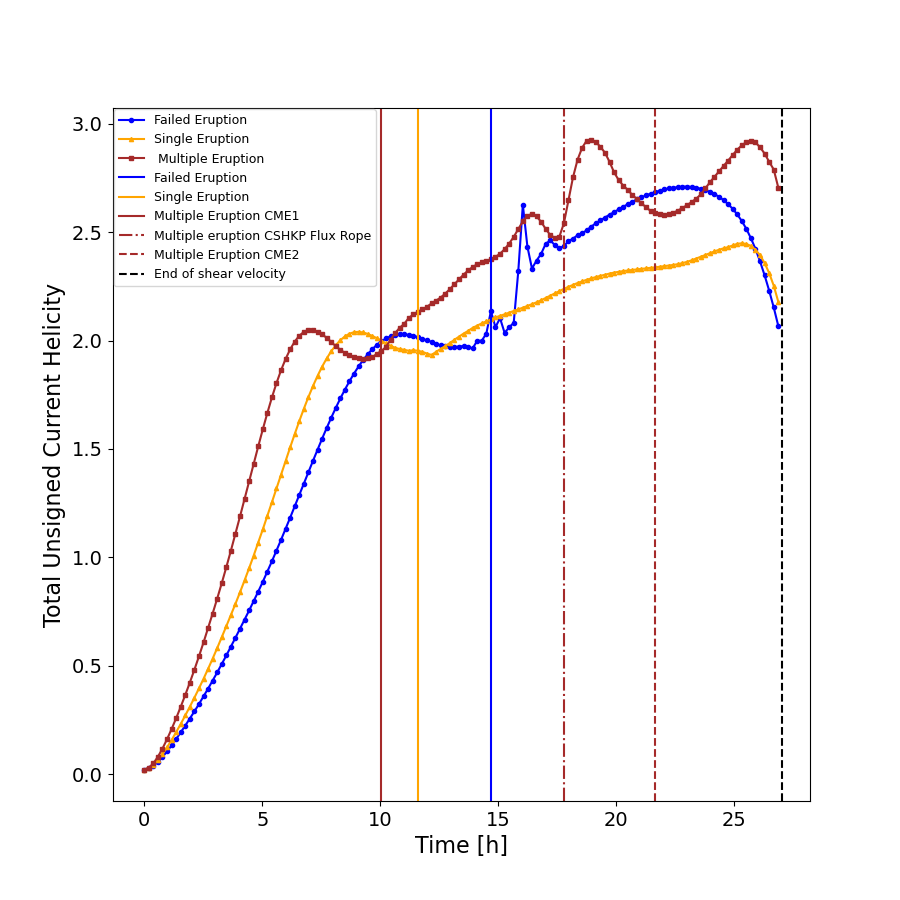}{0.42\textwidth}{ }}

\caption{Temporal evolution of total unsigned current helicity for three scenarios (blue, yellow, and red for failed, single, and multiple eruptions, respectively). The blue, yellow and red vertical lines represent the flux rope formation time for the failed, single and multiple eruption cases. The black dotted vertical line represents the end of the shear. Time is measured from the start of the shear. \label{fig:tuch}}
\end{figure}

\begin{figure}
\gridline{\fig{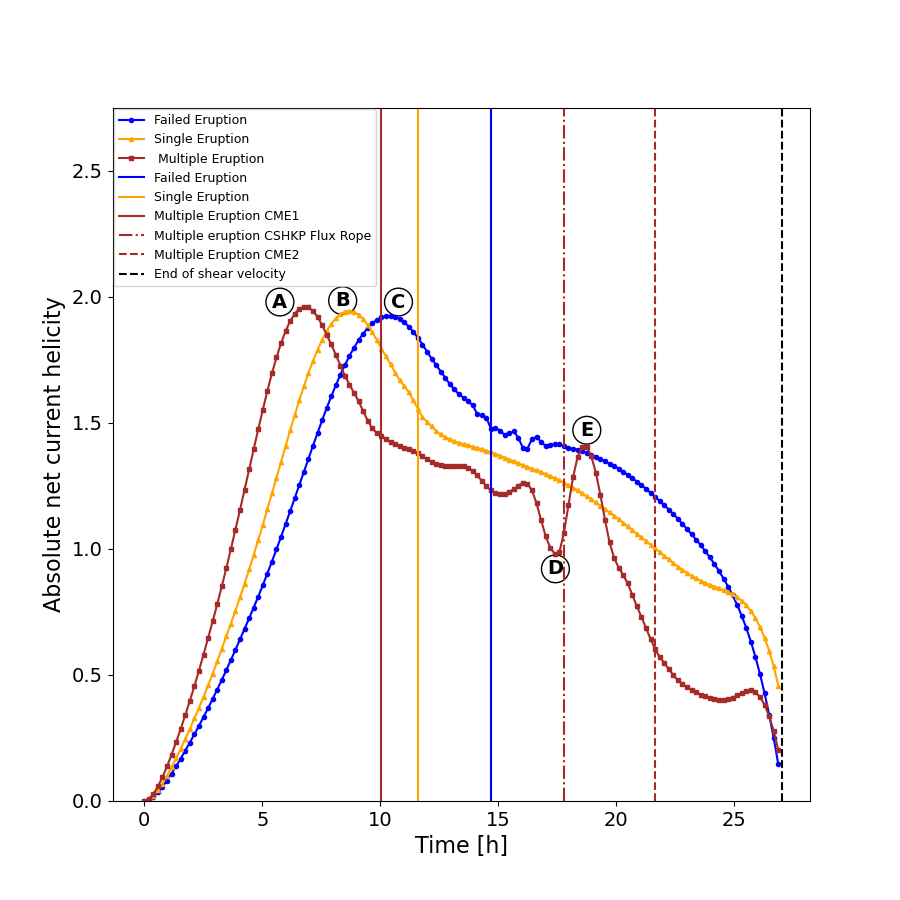}{0.42\textwidth}{ }}

\caption{Temporal evolution of absolute net current helicity for three scenarios (blue, yellow, and red for failed, single, and multiple eruptions, respectively). The blue, yellow and red vertical lines represent the flux rope formation time for the failed, single and multiple eruption cases. The black dotted vertical line represents the end of the shear. Time is measured from the start of the shear. \label{fig:anch}}
\end{figure}

Figures \ref{fig:me}, \ref{fig:tuch} and \ref{fig:anch} depict the temporal evolution of excess magnetic energy, total unsigned current helicity and absolute net current helicity in the arcade system following the application of shearing at the base. Each graph showcases three distinct curves: blue for the failed eruption scenario, yellow for the single eruption scenario, and red for the multiple eruptions scenario. The different vertical lines in each plot represent the time of the flux rope formation for different scenarios: blue solid line for the failed eruption scenario, yellow solid line for the single eruption scenario, red solid line for the first CME flux rope and red dotted line for the second CME flux rope. The black dotted vertical line marks the time of the end of the shear velocity. \par 
In Figure \ref{fig:me}, it is observed that the excess magnetic energy shows a consistent increase over time, reaching its peak before initiating the breakout reconnection in all scenarios. Subsequently, it decreases for the failed and single eruption cases. However, in the case of multiple eruptions, it undergoes a subsequent increase and, after reaching a maximum value, declines following the occurrence of the second breakout reconnection. Notably, the rate of increase in magnetic energy is proportional to the magnitude of the applied shear at the base. When the shear velocity is low, the increase is more gradual, and the maximum value attained is lower. Conversely, an increase in shear velocity at the base corresponds to a higher rate of increase in magnetic energy, accompanied by an increase in the maximum value reached. We also noted that the reduction in magnetic energy following the initial breakout reconnection is more gradual for lower shearing speeds compared to higher shearing speeds.\par

The total unsigned current helicity exhibits a continuous increase over time, as depicted in Figure \ref{fig:tuch}. Prior to the initial breakout reconnection, the rate of increase is more gradual for lower values of applied shear than for higher shear. Importantly, this increase persists even after the formation of the flux rope in all cases. In the failed eruption scenario, a pronounced surge in total unsigned current helicity occurs following flux rope formation. Despite this significant increase, it fails to lead to an eruption. In contrast, for the multiple eruptions case, the helicity value experiences another increase after the formation of the first flux rope. Notably, this rise is more gradual, and it eventually results in another eruption.

The absolute net current helicity profile, as depicted in Figure \ref{fig:anch}, exhibits a continuous increase over time until the occurrence of the first breakout reconnection. Once again, the ascent is more gradual for lower shear at the base compared to higher shear. In Figure \ref{fig:anch}, points A, B, and C denote the maximum values for the multiple, single, and failed eruption cases, respectively. Following the first reconnection, the helicity value diminishes. During this declining phase, flux rope formation is observed in all cases, as reflected in their respective curves. For the failed and single eruption scenarios, this value continues to decrease. However, in the multiple eruptions case, it experiences a resurgence between points D and E. In the case of multiple eruptions, the second maximum value preceding the second eruption is lower than the first maximum value preceding the first eruption. The slope of increase in its value between marks D and E surpasses that of the initial phase for the failed eruption case (between 0 and point C), ultimately resulting in another eruption. We calculated the slope between the first data point and the maximum value (denoted by points A, B and C) for the absolute net current helicity for all the different cases. The value was found to be 0.645, 0.498 and 0.412 ( all in simulation units) for the multiple, single and failed eruption cases, respectively. Failed eruption cases have the lowest value. We also calculate the slope between point D and point E before the second eruption in the case of multiple eruptions cases. This value was found to be 0.801, which is greater than the value for the failed eruption case. Hence, we can conclude that, in addition to the absolute value of the absolute net current helicity, the rate of its increase also plays an important role in influencing the occurrence of the eruption.

\section{Summary}
We perform MHD simulations of a breakout CME using the MPI-AMRVAC code. In this breakout reconnection-driven model, shear is applied at the base of a multipolar magnetic configuration, facilitating the formation of a flux rope that eventually erupts. Initially, we vary the background magnetic field strength to comprehend its impact on the eruption. This investigation aims to understand the influence of global magnetic field changes on localized eruptions. It explores the possibility that a greater number of slow CMEs may occur during weaker solar cycles, such as Solar Cycle 24, which was characterised by reduced heliospheric pressure and weaker polar fields. \cite{Gopalswamy2015} and \cite{petrie2015ApJ...812...74P} argued that such conditions allowed more weaker eruptions from weaker magnetic structures, thereby contributing to the higher CME rate observed during this cycle. In contrast, \cite{wang2014ApJ...784L..27W} suggested that the influence of polar fields on the global CME rate is likely a second-order effect, and that changes in instrumental cadence may explain the apparent increase. Our simulations support the idea that reduced background field strength can be a contributing factor in facilitating eruptions. However, further extensive analysis through simulations along these lines will better establish our insights. \cite{Bemporad2012} also discussed the effect of an increase in the strength of the global dipole magnetic field, where they compared their findings with \citep{Zuccarello_2012}. However, in their cases, the 33 \% change in the strength of the global dipole magnetic field causes significant changes in the magnetic field configuration of the system, which also alters the final result of their simulations. In contrast, our simulations indicate a marginal increase of less than 5\% in the background dipole magnetic field strength, and a consistent magnetic field configuration in different simulation set-up resulted in different outcomes.

In addition to modifying the background magnetic field strength, we also adjusted the amplitude of the shear at the base. This shear influences the helicity injected into the system. Despite the slight variation in shear amplitude, it led to different eruption cases, including failed eruption, single eruption, and multiple eruptions scenarios, supporting the similar results of \citet{dana2020}. This illustrates that the same physical mechanism of eruption behaves differently under subtly varying conditions, thus making the prediction of eruption challenging.

Numerous efforts have been made to predict solar eruptions using data from magnetograms, relying primarily on SHARP parameters. These parameters are typically derived from observations of photospheric vector magnetograms alone. However, through numerical simulations—which provide magnetic field components throughout the entire simulated domain—we were able to calculate these parameters within the solar corona as well. Our analysis focused on examining the evolution of three global magnetic parameters—magnetic energy, total unsigned current helicity, and absolute net current helicity—across different eruption scenarios. Our findings indicate that among these parameters, the time rate of absolute net current helicity can serve as the most effective indicator for distinguishing between various eruption scenarios.
\vspace{5mm}
\section{Acknowledgement} We thank the anonymous referee for the valuable comments that have improved the manuscript. The authors would like to acknowledge the support of the Aryabhatta Research Institute of Observational Sciences (ARIES), Nainital, for facilitation and access to the "Surya" High-Performance Computing (HPC) Facility in the completion of this work. The authors would like to thank the MPI-AMRVAC support team, especially Prof. Roney Keppens, Dr. Chun Xia and Dr. Jack Jenkins, for their help with the MPI-AMRVAC. D.-C. T. thanks the Belgian Federal Science Policy Office (BELSPO) for the provision of financial support in the framework of the PRODEX Programme of the European Space Agency (ESA) under contract number C4000136681. The authors acknowledge the benefits from the discussion of the ISSI-BJ Team ``Solar eruptions: preparing for the next generation multi-wavelength coronagraph". The authors also recognise the open-data source policy of Paraview, Python and yt visualisation tool.
\newpage
\bibliography{sample631}{}
\bibliographystyle{aasjournal}


\end{document}